\documentclass[twocolumn,superscriptaddress,amsmath,amssymb,pra,longbibliography]{revtex4-1}
\usepackage{amssymb,amsmath}
\usepackage{cmap}
\usepackage{graphicx}
\usepackage{bm}% bold math
\usepackage{color}
\usepackage{blindtext}
\usepackage{hyperref}
\usepackage{listings}
\usepackage{booktabs}
\usepackage{multirow}
\usepackage{amsmath}
\usepackage{mathrsfs}
\usepackage{braket}
\usepackage{mathtools}

\usepackage{dsfont}
\usepackage{tabularx}

\begin{document}

\title{Ponderomotive Forces, Stokes Drift, and Momentum \\ in Acoustic and Electromagnetic Waves}

\author{Konstantin Y. Bliokh}
\affiliation{Theoretical Quantum Physics Laboratory, RIKEN Cluster for Pioneering Research, Wako-shi, Saitama 351-0198, Japan}
\author{Yury P. Bliokh}
\affiliation{Physics Department, Technion, Israel Institute of Technology, Haifa 320003, Israel}
\author{Franco Nori}
\affiliation{Theoretical Quantum Physics Laboratory, RIKEN Cluster for Pioneering Research, Wako-shi, Saitama 351-0198, Japan}
\affiliation{Physics Department, University of Michigan, Ann Arbor, Michigan 48109-1040, USA}

%\date{\today}

\begin{abstract}
We reveal universal connections between three important phenomena in classical wave physics: (i) the ponderomotive force acting on the medium particles in an oscillatory wavefield, (ii) the Stokes drift of free medium particles in a wave field, and (iii) the canonical wave momentum in a medium. We analyse these phenomena for (a) longitudinal sound waves in a gas or fluid and (b) transverse electromagnetic waves interacting with electrons in a plasma or metal. In both cases, assuming quasi-monochromatic yet arbitrarily inhomogeneous wavefields, the connections between the ponderomotive force, Stokes drift velocity, kinetic energy, and canonical wave momentum carried by the medium particles are given by the same simple expressions. This sheds light on the nature of these phenomena, their fine interplay, and can be useful for applications to dynamical transport phenomena in various types of waves.    
\end{abstract}

%\keywords{Acoustic force; canonical momentum; acoustic spin; acoustic torque.}

\maketitle

%%%%%%%%%%%%%%%%%%%%%%%%%%%%%%
\section{Introduction}
%%%%%%%%%%%%%%%%%%%%%%%%%%%%%%
Fundamental relations between the velocity ${\bf v}$, momentum ${\bf p}$, and force ${\bf f}$ for a point particle lie at the heart of classical mechanics and Newton's second law: 
\begin{equation}
\label{eq-Newton}
{\bf f} = \frac{d {\bf p}}{d t}\, , \quad {\bf p} = m {\bf v} \, ,
\end{equation}
where $m$ is the mass of the particle. In wave physics, such relations are not so obvious. Indeed, the very definition of the wave momentum causes ongoing debates in acoustics, fluid mechanics, and electrodynamics of continuous media \cite{Peierls_I,Peierls_II,Soper,McIntyre1981,Peskin2010,Stone2002,Brevik1979,Pfeifer2007,Barnett2010_II,Milonni2010,Kemp2011}.
Some approaches associate the wave momentum with the force acting on the medium or probe particles. However, this force depends on the type of wave-matter interaction, properties of the particles, medium inhomogeneities, etc. Radiation forces in acoustic and optical fields are subjects of numerous nontrivial studies \cite{Bruus2012,Meng2019,Toftul2019,Dholakia2020,Ashkin2000,Grier2003,Dienerowitz2008,Bliokh2014,Sukhov2017}.  

In this paper, considering waves in basic continuous media with free particles, we derive simple relations between the ponderomotive force acting on the medium particles in a quasi-monochromatic wave field, the wave energy and canonical wave momentum carried by the particles, and the {\it Stokes drift} \cite{Stokes,Bremer2017,Falkovich_book} of the particles: 
\begin{equation}
\label{eq-force}
\bar{\bf F} = - {\bm \nabla} \bar{W} + \frac{\partial \bar{\bf P}}{\partial t}\, , \quad 
\bar{\bf P} = \rho\, {\bf V}_S \, .
\end{equation}
Here and hereafter, the overbar stands for the cycle-average over oscillations with central frequency $\omega$, $\bar{\bf F}$ is the ponderomotive force density, $W$ is the kinetic energy density of the particles, ${\bf P}$ is the canonical wave momentum density carried by the particles, ${\bf V}_S$ is the velocity of the Stokes drift of the particles, and $\rho$ is the unperturbed mass density of the particles. The first term in the force (\ref{eq-force}) is known as the {gradient force} \cite{Gaponov1958,Gordon1973,Washimi1976}, while the second term can be regarded as the wave analogue of Newton's second law Eq.~(\ref{eq-Newton}). 

Although particular cases of this time-derivative term has appeared in several studies on the electromagnetic wave momentum in dispersive media \cite{Gordon1973,Washimi1976,Milonni2010,Bliokh2017NJP}, and measurements of the radiation force/momentum have also been performed \cite{Jones1978,Campbell2005,Astrath2022}, its general form for waves of different nature and connection with the Stokes drift of the particles have not been described, to the best of our knowledge. Moreover, we show that the wave momentum density $\bar{\bf P}$ is the {\it canonical} field-theory momentum density (or its part carried by the particles), which was properly recognized for structured optical and acoustic wavefields only recently \cite{Berry2009,Bliokh2014NC,Bliokh2014,Bliokh2015PR,Bliokh2017NJP,Shi2019,Toftul2019,Burns2020,Bliokh2022,Bliokh2022_II}. We also emphasize that Eq.~(\ref{eq-force}) describes the force on individual microscopic particles constituting the medium rather than the radiation force on a macroscopic particle immersed in the medium \cite{Bruus2012,Meng2019,Toftul2019,Dholakia2020,Ashkin2000,Grier2003,Dienerowitz2008,Bliokh2014,Sukhov2017}.

Below we consider two basic types of longitudinal and transverse waves in isotropic homogeneous continuous media: (a) sound waves in a fluid or gas and (b) electromagnetic waves in a medium with free electrons (plasma or metal). In both cases Eq.~(\ref{eq-force}) is valid for quasi-monochromatic wavefields with arbitrary spatial inhomogeneities. This is in contrast to some previous studies considering only plane-wave-like fields with well-defined wavevectors \cite{Washimi1976,Milonni2010,Dodin2012}. Furthermore, we consider the effect of losses, which adds a dissipative contribution to the first Eq.~(\ref{eq-force}), also determined by the Stokes drift velocity. Our derivations can be applied to waves of other natures: e.g., the Langmuir plasma waves similar to sound waves \cite{Krall,Bliokh2022_II}.
 
\begin{figure}[t!]
\includegraphics[width=0.95\linewidth]{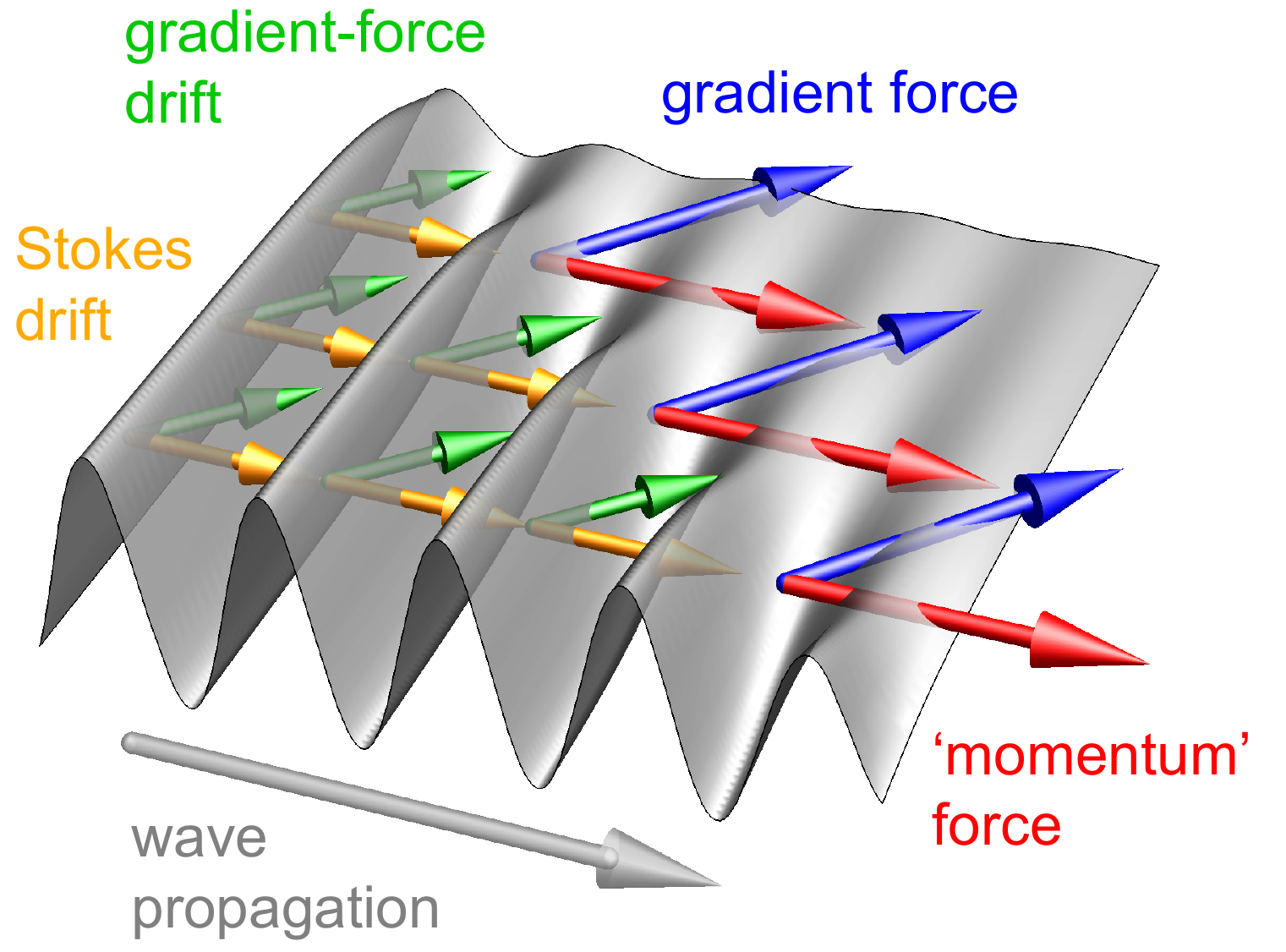}
\caption{Schematics of the propagation of an inhomogeneous sound wave with different directions of the intensity gradient and the local wave vector. As the wave amplitude grows in a given point, the medium particles experience the action of two ponderomotive forces, Eq.~(\ref{eq-force}): (i) the gradient one, proportional to the intensity gradient and (ii) the one due to the time derivative of the wave momentum. These forces accelerate the medium particles and generate the two corresponding drift velocities. The velocity associated with the wave momentum is known as the Stokes drift, Eqs.~(\ref{eq-force}) and (\ref{eq-Stokes}). 
\label{Fig1}}
\end{figure}

Our findings shed light on the nature of the wave momentum in continuous media, poderomotive forces, and their connection with the Stokes drift phenomenon (mostly known in fluid mechanics). For example, the momentum carried by a sound wave packet is produced by the ponderomotive forces which accelerate the medium particles as the wavepacket approaches the given point, so that the particles acquire the Stokes drift velocity (plus an extra velocity from the gradient force) inside the packet, as shown in Fig.~\ref{Fig1}. Since ponderomotive forces, wave momentum, and drifts are highly important for numerous applications, such as optical or acoustic manipulation of particles \cite{Ashkin2000,Grier2003,Dienerowitz2008,Sukhov2017,Meng2019,Dholakia2020}, laser cooling \cite{Stenholm1986,Chu1998,Phillips1998}, optomechanics \cite{Aspelmeyer2014}, and microfluidics \cite{Bruus2012,Ding2013}, our results can have practical implementations.  

%%%%%%%%%%%%%%%%%%%%%%%%%%%%%%%%%%%%%%
%%%%%%%%%%%%%%%%%%%%%%%%%%%%%%%%%%%%%%

%%%%%%%%%%%%%%%%%%%%%%%%%%%%%%
\section{Sound waves in a fluid or gas}
%%%%%%%%%%%%%%%%%%%%%%%%%%%%%%
We start with the equation of motion for particles of a fluid or gas in a linear sound wave \cite{Falkovich_book,LLfluid}:
\begin{equation}
\label{eq-sound}
\rho\, \frac{\partial {\bf v}({\bf r},t)}{\partial t} = - {\bm \nabla} p({\bf r},t) \equiv {\bf F}({\bf r},t) \, ,
\end{equation}
where ${\bf v}$ is the particle Eulerian velocity and $p$ is the pressure. Sound waves are longitudinal: ${\bm \nabla}\times {\bf v} = {\bf 0}$. Since the velocity of a given particle is ${\bf v} = \partial {\bf r}/\partial t$, Eq.~(\ref{eq-sound}) can be regarded as a {\it nonlinear} equation for the particle coordinates ${\bf r}$. Quadratic corrections to the linearized oscillatory motion of the particle are responsible for both the poderomotive force and Stokes drift. Note that we use the Eulerian description of all quantities in a given spatial point, and therefore the local force density ${\bf F}$ is associated with the partial rather than total (material) time derivative of the velocity. 

To find these effects, we use the method of averaging over fast oscillations. We write the particle coordinates as ${\bf r}= {\bf R} + {\bf a}$, where ${\bf R}$ describes the fixed (Eulerian) coordinates, whereas ${\bf a}$ corresponds to the fast oscillatory motion. In the linear approximation, $\rho\, \dfrac{\partial^2 {\bf a}({\bf R},t)}{\partial t^2} = {\bf F}({\bf R},t)$. In the next approximation, the force density in the right-hand side of Eq.~(\ref{eq-sound}) can be written as ${\bf F}({\bf r}) \simeq {\bf F}({\bf R})+ [{\bf a}({\bf R})\cdot {\bm \nabla} ]\, {\bf F}({\bf R})$.   
We now introduce the complex-amplitude representation for all linear quasi-monochromatic wave fields: 
${\bf v}({\bf R},t) = {\rm Re}\!\left[ {\bm{\mathsf v}}({\bf R},t) e^{-i\omega t} \right]$, ${\bf F}({\bf R},t) = {\rm Re}\!\left[ {\bm{\mathsf F}}({\bf R},t) e^{-i\omega t} \right]$, ${\bf a}({\bf R},t) = {\rm Re}\!\left[ {\bm{\mathsf a}}({\bf R},t) e^{-i\omega t} \right]$, etc. Substituting this representation into the above expansion for the force and performing the time averaging over one period of fast oscillations with frequency $\omega$, we obtain the time-averaged quadratic ponderomotive force density:
\begin{equation}
\label{eq-sound-force1}
\bar{\bf F} = \frac{1}{2}\,{\rm Re}\!\left[({\bm{\mathsf a}}^*\cdot  {\bm \nabla}) {\bm{\mathsf F}}\right] .
%= \frac{\rho}{2}{\rm Re}\!\left[({\bf a}^*\cdot \nabla) \frac{\partial{\bf v}}{\partial t}\right] \, ,
\end{equation}
Using the relations ${\bm{\mathsf v}} = \partial {\bm{\mathsf a}}/\partial t -i\omega\, {\bm{\mathsf a}}$, ${\bm{\mathsf F}} = \rho\, \partial {\bm{\mathsf v}}/\partial t -i\rho\,\omega\, {\bm{\mathsf v}}$, ${\bm \nabla}\times {\bm{\mathsf v}} = {\bf 0}$ and some algebra, Eq.~(\ref{eq-sound-force1}) can be written as
\begin{equation}
\label{eq-sound-force2}
\bar{\bf F} = -\frac{\rho}{4}\, {\bm \nabla} |{\bm{\mathsf v}}|^2 + \frac{\rho}{2}\, \frac{\partial}{\partial t}{\rm Re}\!\left[({\bm{\mathsf a}}^*\cdot {\bm\nabla}) {\bm{\mathsf v}}\right] .
\end{equation}

Equation~(\ref{eq-sound-force2}) has the form of the first Eq.~(\ref{eq-force}) if we note that the time-averaged kinetic energy density is $\bar{W} = \rho\, |{\bm{\mathsf v}}|^2 / 4$ and the canonical momentum density of sound waves is 
$\bar{\bf P} =\dfrac{\rho}{2}\,{\rm Re}\!\left[({\bm{\mathsf a}}^*\cdot {\bm\nabla}) {\bm{\mathsf v}}\right]$. The latter equation yields 
\begin{equation}
\label{eq-sound-momentum}
\bar{\bf P} =\dfrac{\rho}{2\omega}\,{\rm Im}\!\left[{\bm{\mathsf v}}^*\cdot ({\bm \nabla}) {\bm{\mathsf v}}\right]
\end{equation}
for monochromatic fields which agrees with recent derivations \cite{Shi2019,Burns2020,Bliokh2022,Bliokh2022_II}. 

The Stokes drift \cite{Stokes,Bremer2017,Falkovich_book} is known to appear from the difference between the Lagrangian and Eulerian velocities of the particle. It can be derived from the expansion of the velocity similarly to the force above: ${\bf v}({\bf r}) \simeq {\bf v}({\bf R})+ [{\bf a}({\bf R})\cdot  {\bm \nabla} ]\, {\bf v}({\bf R})$. Performing the time averaging for quasi-monochromatic fields yields the Stokes drift velocity 
\begin{equation}
\label{eq-Stokes}
{\bf V}_S \equiv \bar{\bf v} = \frac{1}{2}{\rm Re}\! \left[ ({\bm{\mathsf a}}^*\cdot {\bm \nabla}) {\bm{\mathsf v}} \right].
%= \frac{1}{2}{\rm Re}\! \left[ {\bf a}^*\cdot (\nabla) {\bf v} \right] ,
\end{equation}
This yields the second Eq.~(\ref{eq-force}). 

For a plane sound wave with wavevector ${\bf k}$, ${\bm \nabla} \to i {\bf k}$, the canonical momentum density and the Stokes drift velocity become $\bar{\bf P} = \rho\, \bar{\bf V}_S = 2\, {\bf k}\, \bar{W} /\omega$. Here the factor of 2 is related to the fact that the total sound-wave energy, consisting of the kinetic and potential parts, is $2\bar{W}$ \cite{Falkovich_book,LLfluid,Burns2020}.

%%%%%%%%%%%%%%%%%%%%%%%%%%%%%%
\section{Electromagnetic waves interacting with electrons in a plasma or metal}
%%%%%%%%%%%%%%%%%%%%%%%%%%%%%%
We now consider a different kind of waves: electromagnetic waves in a medium with free electrons, i.e., a plasma or metal \cite{Krall,Maier_book}. The equation of motion for electrons in an electromagnetic wave involves the Lorentz force:   
\begin{equation}
\label{eq-EM}
\rho\, \frac{\partial {\bf v}({\bf r},t)}{\partial t} = en\!\left[ {\bf E}({\bf r},t) + \frac{{\bf v}({\bf r},t) \times {\bf H}({\bf r},t)}{c} \right], %\equiv {\bf F}({\bf r},t), 
%\\ \rho\, \frac{\partial {\bf v}({\bf r},t)}{\partial t} = - {\bm \nabla} p({\bf r},t) \equiv {\bf F}({\bf r},t) \, ,
\end{equation}
where ${\bf E}$ and ${\bf H}$ are the electric and magnetic fields, which obey Maxwell's equations, $e<0$ is the electron charge, $n$ is the concentration of electrons, $c$ is the speed of light, and we use Gaussian units.
The heavy ion background in plasma neutralizes the total electric charge of the electrons and is assumed to be motionless in this approximation.
Note that $\rho = m n$, where $m$ is the electron mass, so that Eq.~(\ref{eq-EM}) is equivalent to the single-electron equation of motion. We write it in terms of volume densities to derive the densities of force, momentum, etc. akin to the sound-wave case.
In contrast to sound waves, electromagnetic waves are transverse: ${\bm \nabla}\cdot {\bf E} = {\bm \nabla}\cdot {\bf H} = 0$.

To derive the quadratic ponderomotive force and Stokes drift of electrons we follow the same approach as for sound waves. The only difference now is that the Lorentz force from the magnetic field ${\bf H}$ is {\it quadratic} in the wave amplitude and should be neglected in the linear approximation. Therefore, linear oscillations of the electrons are described by the equation 
$\rho\, \dfrac{\partial^2 {\bf a}({\bf R},t)}{\partial t^2} = en\,{\bf E}({\bf R},t)$, whereas the quadratic time-averaged force density for a quasi-monochromatic wavefield is given by
\begin{equation}
\label{eq-EM-force1}
\bar{\bf F} = \frac{en}{2}{\rm Re}\!\left[({\bm{\mathsf a}}^*\cdot {\bm \nabla}) {\bm{\mathsf E}}\right]
+ \frac{en}{2c}{\rm Re}\!\left({\bm{\mathsf v}}^* \times {\bm{\mathsf H}}\right) .
\end{equation}
Notably, introducing the dipole moment ${\bf d} = e\, {\bf a}$, the force (\ref{eq-EM-force1}) becomes equivalent to the electric-dipole part of the cycle-averaged Einstein-Laub force \cite{Einstein1908,Mansuripur2014,Webb2016}. Thus, the two descriptions, in terms of individual charges and in terms of wave-induced dipoles, are equivalent to each other.
Using the relations ${\bm{\mathsf v}} = \partial {\bm{\mathsf a}}/\partial t -i\omega\, {\bm{\mathsf a}}$, $en{\bm{\mathsf E}} = \rho\, \partial {\bm{\mathsf v}}/\partial t -i\rho\,\omega\, {\bm{\mathsf v}}$, 
%${\bm \nabla}\cdot {\bm{\mathsf v}} = 0$, 
the Maxwell equation $c {\bm \nabla} \times {\bm{\mathsf E}} = - {\partial{\bm{\mathsf H}}}/{\partial t} + i\omega {\bm{\mathsf H}}$ (yielding $en{\bm{\mathsf H}} = - \rho c {\bm \nabla} \times {\bm{\mathsf v}}$), and some algebra, Eq.~(\ref{eq-EM-force1}) can be transformed into exactly the same form as Eq.~(\ref{eq-sound-force2}):
\begin{equation}
\label{eq-EM-force2}
\bar{\bf F} = -\frac{\rho}{4}\, {\bm \nabla} |{\bm{\mathsf v}}|^2 + \frac{\rho}{2}\, \frac{\partial}{\partial t}{\rm Re}\!\left[({\bm{\mathsf a}}^*\cdot {\bm\nabla}) {\bm{\mathsf v}}\right] .
\end{equation}
Obviously, $\bar{W} = \rho\, |{\bm{\mathsf v}}|^2 / 4$ is the time-averaged kinetic energy density for electrons, and $\bar{\bf P} ={\rho}{\bf V}_S$ is their momentum density associated with the Stokes drift velocity (\ref{eq-Stokes}). Thus, Eq.~(\ref{eq-EM-force2}) has the desired form of the first Eq.~(\ref{eq-force}). 

In contrast to longitudinal sound waves, the Stokes drift velocity and time-averaged momentum of electrons {\it vanish} in a plane electromagnetic wave: $\bar{\bf P} = {\bf V}_S ={\bf 0}$, because of its transverse character: ${\bf k}\cdot {\bm{\mathsf a}} =0$. Nonetheless, these quantities are generally nonzero and play important roles in structured electromagnetic waves. For monochromatic fields, using the equations of motion, the electron wave momentum can be written as \cite{Bliokh2022_II}
\begin{equation}
\label{eq-EM-momentum}
\bar{\bf P} =-\frac{\rho}{4\omega}{\bm \nabla}\!\times {\rm Im}\!\left({\bm{\mathsf v}}^*\!\times {\bm{\mathsf v}}\right)
= -\frac{\omega_p^2}{16\pi\omega^3}{\bm \nabla}\!\times {\rm Im}\!\left({\bm{\mathsf E}}^*\!\times {\bm{\mathsf E}}\right) ,
\end{equation}
where $\omega_p=\sqrt{4\pi n^2e^2/\rho}$ is the electron plasma frequency \cite{Krall,Maier_book}. By adding this contribution to the Poynting momentum density carried by the electromagnetic field, $\bar{\bf P}_{\rm field} = \dfrac{1}{8\pi c} {\rm Re}({\bm{\mathsf E}}^*\!\times {\bm{\mathsf H}})$, one can derive the canonical (i.e., Minkowski with proper dispersive corrections) momentum density for monochromatic electromagnetic wave in a medium with permittivity $\varepsilon (\omega) = 1 - \omega_p^2 / \omega^2$ \cite{Philbin2011,Philbin2012,Bliokh2017NJP}.

Thus, the Stokes drift and the corresponding momentum of electrons provide an important contribution to the total momentum of an electromagnetic wave in a plasma or metal. This contribution vanishes for a plane wave because the dispersive correction to the wave momentum from the frequency-dependent permittivity $\varepsilon$ \cite{Nelson1991,Philbin2011,Philbin2012,Bliokh2017NJP} exactly cancels the difference between the Minkowski and Abraham wave momenta, $\bar{\bf P}_M = \varepsilon\, \bar{\bf P}_{\rm field}$ and $\bar{\bf P}_A = \bar{\bf P}_{\rm field}$ \cite{Pfeifer2007,Barnett2010,Milonni2010,Kemp2011}. In the interpretation of Ref.~\cite{Partanen2017} this is explained by the fact that electromagnetic waves in a plasma-like medium do not excite mass density waves.

%%%%%%%%%%%%%%%%%%%%%%%%%%%%%%
%\subsection{Effects of dissipation}
%%%%%%%%%%%%%%%%%%%%%%%%%%%%%%
Consider now effects of the medium absorption on the ponderomotive force. Absorption can be introduced via an effective friction force ${\bf F}_{\rm diss} = -\gamma\, \rho\, {\bf v}$ in the right-hand side of the equation of motion (\ref{eq-sound}) or (\ref{eq-EM}). In the electromagnetic-wave case, this results in the complex permittivity of plasma or metal: $\varepsilon (\omega) = 1 - \dfrac{\omega_p^2}{\omega (\omega + i \gamma)}$ \cite{Krall,Maier_book}. Substituting the dissipative force into Eqs.~(\ref{eq-sound-force1}) or (\ref{eq-EM-force1}) and using Eq.~(\ref{eq-Stokes}), we obtain the dissipative ponderomotive force proportional to the Stokes drift velocity: 
\begin{equation}
\label{eq-diss}
\bar{\bf F}_{\rm diss} =-\gamma\, \rho\, {\bf V}_S\, .
\end{equation}
This force is neither a gradient force nor a derivative of the wave momentum, and it should be considered as the third contribution to the first Eq.~(\ref{eq-force}). 

%%%%%%%%%%%%%%%%%%%%%%%%%%%%%%
\section{Conclusions}
%%%%%%%%%%%%%%%%%%%%%%%%%%%%%%
We have examined ponderomotive forces produced by quasi-monochromatic wave fields in continuous media with free particles. We considered both longitudinal sound waves and transverse electromagnetic waves. In both cases the ponderomotive force has the same form (\ref{eq-force}) consisting of two contributions. The first contribution is the well-known gradient force proportional to the gradient of the kinetic energy density of the medium particles. The second contribution is given by the time derivative of the particle contribution to the canonical momentum density in the wave. We have shown that this momentum density is associated with the Stokes drift of the particles. In dissipative media, the ponderomotive force has the third, dissipative contribution (\ref{eq-diss}), which is also determined by the Stokes drift velocity. Thus, our results illuminate close interrelations between the poderomotive force, wave momentum, and Stokes drift. 

For sound waves in a fluid or gas, the Stokes drift and the corresponding particle momentum provide the total canonical momentum of the wave. For electromagnetic waves, the Stokes-drift momentum of electrons is only a part of the total wave momentum; the remaining part is given by the electromagnetic field momentum. Notably, the Stokes-drift momentum provides an important contribution for the resolution of the Abraham-Minkowski dilemma in a metal or plasma \cite{Bliokh2017NJP,Bliokh2022_II}. 

When a wavepacket or another inhomogeneous wave propagates in a medium, the ponderomotive forces accelerate the medium particles near the front edge of the wavepacket and produce the corresponding particle drift inside the wavepacket, Fig.~\ref{Fig1}. Thus, the gradient and momentum-related forces in Eq.~(\ref{eq-force}) generate two contributions to the total drift and momentum of particles. This explains (at least partly) why the Stokes drift is difficult to observe in its pure form \cite{Bremer2017}: it is always accompanied by another drift from the gradient force. However, the drift from the gradient force cannot be associated with the wave momentum: it can appear even in a standing wave without any propagation. 

\vspace*{0.3cm}
\begin{acknowledgments}
%%%%%%%%%%%%%%%%%%%%%%%%%%%%%%
%\section*{ACKNOWLEDGMENTS}
%%%%%%%%%%%%%%%%%%%%%%%%%%%%%%
This work was partially supported by Nippon Telegraph and Telephone Corporation (NTT) Research;
the Japan Science and Technology Agency (JST) via
the Quantum Leap Flagship Program (Q-LEAP) and
the Moonshot R\&D Grant No.~JPMJMS2061;
the Japan Society for the Promotion of Science (JSPS)
via the Grants-in-Aid for Scientific Research (KAKENHI) Grant No.~JP20H00134;
the Army Research Office (ARO) (Grant No.~W911NF-18-1-0358),
the Asian Office of Aerospace Research and Development (AOARD) (Grant No.~FA2386-20-1-4069); and
the Foundational Questions Institute Fund (FQXi) (Grant No.~FQXi-IAF19-06).
\end{acknowledgments}

%\newpage

%\pagebreak

\bibliography{References_Stokes}

\begin{thebibliography}{54}%
\makeatletter
\providecommand \@ifxundefined [1]{%
 \@ifx{#1\undefined}
}%
\providecommand \@ifnum [1]{%
 \ifnum #1\expandafter \@firstoftwo
 \else \expandafter \@secondoftwo
 \fi
}%
\providecommand \@ifx [1]{%
 \ifx #1\expandafter \@firstoftwo
 \else \expandafter \@secondoftwo
 \fi
}%
\providecommand \natexlab [1]{#1}%
\providecommand \enquote  [1]{``#1''}%
\providecommand \bibnamefont  [1]{#1}%
\providecommand \bibfnamefont [1]{#1}%
\providecommand \citenamefont [1]{#1}%
\providecommand \href@noop [0]{\@secondoftwo}%
\providecommand \href [0]{\begingroup \@sanitize@url \@href}%
\providecommand \@href[1]{\@@startlink{#1}\@@href}%
\providecommand \@@href[1]{\endgroup#1\@@endlink}%
\providecommand \@sanitize@url [0]{\catcode `\\12\catcode `\$12\catcode
  `\&12\catcode `\#12\catcode `\^12\catcode `\_12\catcode `\%12\relax}%
\providecommand \@@startlink[1]{}%
\providecommand \@@endlink[0]{}%
\providecommand \url  [0]{\begingroup\@sanitize@url \@url }%
\providecommand \@url [1]{\endgroup\@href {#1}{\urlprefix }}%
\providecommand \urlprefix  [0]{URL }%
\providecommand \Eprint [0]{\href }%
\providecommand \doibase [0]{http://dx.doi.org/}%
\providecommand \selectlanguage [0]{\@gobble}%
\providecommand \bibinfo  [0]{\@secondoftwo}%
\providecommand \bibfield  [0]{\@secondoftwo}%
\providecommand \translation [1]{[#1]}%
\providecommand \BibitemOpen [0]{}%
\providecommand \bibitemStop [0]{}%
\providecommand \bibitemNoStop [0]{.\EOS\space}%
\providecommand \EOS [0]{\spacefactor3000\relax}%
\providecommand \BibitemShut  [1]{\csname bibitem#1\endcsname}%
\let\auto@bib@innerbib\@empty
%</preamble>
\bibitem [{\citenamefont {Peierls}(1979)}]{Peierls_I}%
  \BibitemOpen
  \bibfield  {author} {\bibinfo {author} {\bibfnamefont {R.}~\bibnamefont
  {Peierls}},\ }\href@noop {} {\emph {\bibinfo {title} {Surprises in
  Theoretical Physics}}}\ (\bibinfo  {publisher} {Princeton University Press,
  Princeton},\ \bibinfo {year} {1979})\BibitemShut {NoStop}%
\bibitem [{\citenamefont {Peierls}(1991)}]{Peierls_II}%
  \BibitemOpen
  \bibfield  {author} {\bibinfo {author} {\bibfnamefont {R.}~\bibnamefont
  {Peierls}},\ }\href@noop {} {\emph {\bibinfo {title} {More Surprises in
  Theoretical Physics}}}\ (\bibinfo  {publisher} {Princeton University Press,
  Princeton},\ \bibinfo {year} {1991})\BibitemShut {NoStop}%
\bibitem [{\citenamefont {Soper}(1976)}]{Soper}%
  \BibitemOpen
  \bibfield  {author} {\bibinfo {author} {\bibfnamefont {D.~E.}\ \bibnamefont
  {Soper}},\ }\href@noop {} {\emph {\bibinfo {title} {Classical Field
  Theory}}}\ (\bibinfo  {publisher} {{Wiley, New York}},\ \bibinfo {year}
  {1976})\BibitemShut {NoStop}%
\bibitem [{\citenamefont {McIntyre}(1981)}]{McIntyre1981}%
  \BibitemOpen
  \bibfield  {author} {\bibinfo {author} {\bibfnamefont {M.~E.}\ \bibnamefont
  {McIntyre}},\ }\bibfield  {title} {\enquote {\bibinfo {title} {On the `wave
  momentum' myth},}\ }\href@noop {} {\bibfield  {journal} {\bibinfo  {journal}
  {J. Fluid Mech.}\ }\textbf {\bibinfo {volume} {106}},\ \bibinfo {pages}
  {331--347} (\bibinfo {year} {1981})}\BibitemShut {NoStop}%
\bibitem [{\citenamefont {Peskin}(2010)}]{Peskin2010}%
  \BibitemOpen
  \bibfield  {author} {\bibinfo {author} {\bibfnamefont {C.~S.}\ \bibnamefont
  {Peskin}},\ }\bibfield  {title} {\enquote {\bibinfo {title} {Wave
  momentum},}\ }\href@noop {} {\bibfield  {journal} {\bibinfo  {journal} {{The
  Silver Dialogues, New York Univ.}}\ } (\bibinfo {year} {2010})}\BibitemShut
  {NoStop}%
\bibitem [{\citenamefont {Stone}(2002)}]{Stone2002}%
  \BibitemOpen
  \bibfield  {author} {\bibinfo {author} {\bibfnamefont {M.}~\bibnamefont
  {Stone}},\ }\bibfield  {title} {\enquote {\bibinfo {title} {Phonons and
  forces: Momentum versus pseudomomentum in moving fluids},}\ }in\ \href@noop
  {} {\emph {\bibinfo {booktitle} {Artificial Black Holes}}},\ \bibinfo
  {editor} {edited by\ \bibinfo {editor} {\bibfnamefont {M.}~\bibnamefont
  {Novello}}, \bibinfo {editor} {\bibfnamefont {M.}~\bibnamefont {Visser}}, \
  and\ \bibinfo {editor} {\bibfnamefont {G.}~\bibnamefont {Volovik}}}\
  (\bibinfo  {publisher} {World Scientific, Singapore},\ \bibinfo {year}
  {2002})\BibitemShut {NoStop}%
\bibitem [{\citenamefont {Brevik}(1979)}]{Brevik1979}%
  \BibitemOpen
  \bibfield  {author} {\bibinfo {author} {\bibfnamefont {I.}~\bibnamefont
  {Brevik}},\ }\bibfield  {title} {\enquote {\bibinfo {title} {Experiments in
  phenomenological electrodynamics and the electromagnetic energy-momentum
  tensor},}\ }\href@noop {} {\bibfield  {journal} {\bibinfo  {journal} {Phys.
  Rep.}\ }\textbf {\bibinfo {volume} {52}},\ \bibinfo {pages} {133--201}
  (\bibinfo {year} {1979})}\BibitemShut {NoStop}%
\bibitem [{\citenamefont {Pfeifer}\ \emph {et~al.}(2007)\citenamefont
  {Pfeifer}, \citenamefont {Nieminen}, \citenamefont {Heckenberg},\ and\
  \citenamefont {Rubinsztein-Dunlop}}]{Pfeifer2007}%
  \BibitemOpen
  \bibfield  {author} {\bibinfo {author} {\bibfnamefont {R.~N.~C.}\
  \bibnamefont {Pfeifer}}, \bibinfo {author} {\bibfnamefont {T.~A.}\
  \bibnamefont {Nieminen}}, \bibinfo {author} {\bibfnamefont {N.~R.}\
  \bibnamefont {Heckenberg}}, \ and\ \bibinfo {author} {\bibfnamefont
  {H.}~\bibnamefont {Rubinsztein-Dunlop}},\ }\bibfield  {title} {\enquote
  {\bibinfo {title} {Colloquium: Momentum of an electromagnetic wave in
  dielectric media},}\ }\href@noop {} {\bibfield  {journal} {\bibinfo
  {journal} {Rev. Mod. Phys.}\ }\textbf {\bibinfo {volume} {79}},\ \bibinfo
  {pages} {1197--1216} (\bibinfo {year} {2007})}\BibitemShut {NoStop}%
\bibitem [{\citenamefont {Barnett}\ and\ \citenamefont
  {Loudon}(2010)}]{Barnett2010_II}%
  \BibitemOpen
  \bibfield  {author} {\bibinfo {author} {\bibfnamefont {S.~M.}\ \bibnamefont
  {Barnett}}\ and\ \bibinfo {author} {\bibfnamefont {R.}~\bibnamefont
  {Loudon}},\ }\bibfield  {title} {\enquote {\bibinfo {title} {The enigma of
  optical momentum in a medium},}\ }\href@noop {} {\bibfield  {journal}
  {\bibinfo  {journal} {Trans. R. Soc. A.}\ }\textbf {\bibinfo {volume}
  {368}},\ \bibinfo {pages} {927--939} (\bibinfo {year} {2010})}\BibitemShut
  {NoStop}%
\bibitem [{\citenamefont {Milonni}\ and\ \citenamefont
  {Boyd}(2010)}]{Milonni2010}%
  \BibitemOpen
  \bibfield  {author} {\bibinfo {author} {\bibfnamefont {Peter~W.}\
  \bibnamefont {Milonni}}\ and\ \bibinfo {author} {\bibfnamefont {Robert~W.}\
  \bibnamefont {Boyd}},\ }\bibfield  {title} {\enquote {\bibinfo {title}
  {Momentum of light in a dielectric medium},}\ }\href@noop {} {\bibfield
  {journal} {\bibinfo  {journal} {Adv. Opt. Photon.}\ }\textbf {\bibinfo
  {volume} {2}},\ \bibinfo {pages} {519--553} (\bibinfo {year}
  {2010})}\BibitemShut {NoStop}%
\bibitem [{\citenamefont {Kemp}(2011)}]{Kemp2011}%
  \BibitemOpen
  \bibfield  {author} {\bibinfo {author} {\bibfnamefont {B.~A.}\ \bibnamefont
  {Kemp}},\ }\bibfield  {title} {\enquote {\bibinfo {title} {{Resolution of the
  Abraham-Minkowski debate: Implications for the electromagnetic wave theory of
  light in matter}},}\ }\href@noop {} {\bibfield  {journal} {\bibinfo
  {journal} {J. Appl. Phys.}\ }\textbf {\bibinfo {volume} {109}},\ \bibinfo
  {pages} {111101} (\bibinfo {year} {2011})}\BibitemShut {NoStop}%
\bibitem [{\citenamefont {Bruus}(2012)}]{Bruus2012}%
  \BibitemOpen
  \bibfield  {author} {\bibinfo {author} {\bibfnamefont {H.}~\bibnamefont
  {Bruus}},\ }\bibfield  {title} {\enquote {\bibinfo {title} {{Acoustofluidics
  7: The acoustic radiation force on small particles}},}\ }\href@noop {}
  {\bibfield  {journal} {\bibinfo  {journal} {Lab Chip.}\ }\textbf {\bibinfo
  {volume} {12}},\ \bibinfo {pages} {1014} (\bibinfo {year}
  {2012})}\BibitemShut {NoStop}%
\bibitem [{\citenamefont {Meng}\ \emph {et~al.}(2019)\citenamefont {Meng},
  \citenamefont {Cai}, \citenamefont {Li}, \citenamefont {Zhou}, \citenamefont
  {Niu},\ and\ \citenamefont {Zheng}}]{Meng2019}%
  \BibitemOpen
  \bibfield  {author} {\bibinfo {author} {\bibfnamefont {L.}~\bibnamefont
  {Meng}}, \bibinfo {author} {\bibfnamefont {F.}~\bibnamefont {Cai}}, \bibinfo
  {author} {\bibfnamefont {F.}~\bibnamefont {Li}}, \bibinfo {author}
  {\bibfnamefont {W.}~\bibnamefont {Zhou}}, \bibinfo {author} {\bibfnamefont
  {L.}~\bibnamefont {Niu}}, \ and\ \bibinfo {author} {\bibfnamefont
  {H.}~\bibnamefont {Zheng}},\ }\bibfield  {title} {\enquote {\bibinfo {title}
  {Acoustic tweezers},}\ }\href@noop {} {\bibfield  {journal} {\bibinfo
  {journal} {J. Phys. D: Appl. Phys.}\ }\textbf {\bibinfo {volume} {52}},\
  \bibinfo {pages} {273001} (\bibinfo {year} {2019})}\BibitemShut {NoStop}%
\bibitem [{\citenamefont {Toftul}\ \emph {et~al.}(2019)\citenamefont {Toftul},
  \citenamefont {Bliokh}, \citenamefont {Petrov},\ and\ \citenamefont
  {Nori}}]{Toftul2019}%
  \BibitemOpen
  \bibfield  {author} {\bibinfo {author} {\bibfnamefont {I.~D.}\ \bibnamefont
  {Toftul}}, \bibinfo {author} {\bibfnamefont {K.~Y.}\ \bibnamefont {Bliokh}},
  \bibinfo {author} {\bibfnamefont {M.~I.}\ \bibnamefont {Petrov}}, \ and\
  \bibinfo {author} {\bibfnamefont {F.}~\bibnamefont {Nori}},\ }\bibfield
  {title} {\enquote {\bibinfo {title} {Acoustic radiation force and torque on
  small particles as measures of the canonical momentum and spin densities},}\
  }\href@noop {} {\bibfield  {journal} {\bibinfo  {journal} {Phys. Rev. Lett.}\
  }\textbf {\bibinfo {volume} {123}},\ \bibinfo {pages} {183901} (\bibinfo
  {year} {2019})}\BibitemShut {NoStop}%
\bibitem [{\citenamefont {Dholakia}\ \emph {et~al.}(2020)\citenamefont
  {Dholakia}, \citenamefont {Drinkwater},\ and\ \citenamefont
  {Ritsch-Marte}}]{Dholakia2020}%
  \BibitemOpen
  \bibfield  {author} {\bibinfo {author} {\bibfnamefont {K.}~\bibnamefont
  {Dholakia}}, \bibinfo {author} {\bibfnamefont {B.~W.}\ \bibnamefont
  {Drinkwater}}, \ and\ \bibinfo {author} {\bibfnamefont {M.}~\bibnamefont
  {Ritsch-Marte}},\ }\bibfield  {title} {\enquote {\bibinfo {title} {Comparing
  acoustic and optical forces for biomedical research},}\ }\href@noop {}
  {\bibfield  {journal} {\bibinfo  {journal} {Nat. Rev. Phys.}\ }\textbf
  {\bibinfo {volume} {2}},\ \bibinfo {pages} {480--491} (\bibinfo {year}
  {2020})}\BibitemShut {NoStop}%
\bibitem [{\citenamefont {Ashkin}(2000)}]{Ashkin2000}%
  \BibitemOpen
  \bibfield  {author} {\bibinfo {author} {\bibfnamefont {A.}~\bibnamefont
  {Ashkin}},\ }\bibfield  {title} {\enquote {\bibinfo {title} {History of
  optical trapping and manipulation of small-neutral particle, atoms, and
  molecules},}\ }\href@noop {} {\bibfield  {journal} {\bibinfo  {journal} {IEEE
  J. Sel. Top. Quant. Electron.}\ }\textbf {\bibinfo {volume} {6}},\ \bibinfo
  {pages} {841} (\bibinfo {year} {2000})}\BibitemShut {NoStop}%
\bibitem [{\citenamefont {Grier}(2003)}]{Grier2003}%
  \BibitemOpen
  \bibfield  {author} {\bibinfo {author} {\bibfnamefont {D.~G.}\ \bibnamefont
  {Grier}},\ }\bibfield  {title} {\enquote {\bibinfo {title} {A revolution in
  optical manipulation},}\ }\href@noop {} {\bibfield  {journal} {\bibinfo
  {journal} {Nature}\ }\textbf {\bibinfo {volume} {424}},\ \bibinfo {pages}
  {810} (\bibinfo {year} {2003})}\BibitemShut {NoStop}%
\bibitem [{\citenamefont {Dienerowitz}\ \emph {et~al.}(2008)\citenamefont
  {Dienerowitz}, \citenamefont {Mazilu},\ and\ \citenamefont
  {Dholakia}}]{Dienerowitz2008}%
  \BibitemOpen
  \bibfield  {author} {\bibinfo {author} {\bibfnamefont {M.}~\bibnamefont
  {Dienerowitz}}, \bibinfo {author} {\bibfnamefont {M.}~\bibnamefont {Mazilu}},
  \ and\ \bibinfo {author} {\bibfnamefont {K.}~\bibnamefont {Dholakia}},\
  }\bibfield  {title} {\enquote {\bibinfo {title} {Optical manipulation of
  nanoparticles: a review},}\ }\href@noop {} {\bibfield  {journal} {\bibinfo
  {journal} {J. Nanophoton.}\ }\textbf {\bibinfo {volume} {2}},\ \bibinfo
  {pages} {021875} (\bibinfo {year} {2008})}\BibitemShut {NoStop}%
\bibitem [{\citenamefont {Bliokh}\ \emph
  {et~al.}(2014{\natexlab{a}})\citenamefont {Bliokh}, \citenamefont {Kivshar},\
  and\ \citenamefont {Nori}}]{Bliokh2014}%
  \BibitemOpen
  \bibfield  {author} {\bibinfo {author} {\bibfnamefont {K.~Y.}\ \bibnamefont
  {Bliokh}}, \bibinfo {author} {\bibfnamefont {Y.~S.}\ \bibnamefont {Kivshar}},
  \ and\ \bibinfo {author} {\bibfnamefont {F.}~\bibnamefont {Nori}},\
  }\bibfield  {title} {\enquote {\bibinfo {title} {Magnetoelectric effects in
  local light-matter interactions},}\ }\href@noop {} {\bibfield  {journal}
  {\bibinfo  {journal} {Phys. Rev. Lett.}\ }\textbf {\bibinfo {volume} {113}},\
  \bibinfo {pages} {033601} (\bibinfo {year} {2014}{\natexlab{a}})}\BibitemShut
  {NoStop}%
\bibitem [{\citenamefont {Sukhov}\ and\ \citenamefont
  {Dogariu}(2017)}]{Sukhov2017}%
  \BibitemOpen
  \bibfield  {author} {\bibinfo {author} {\bibfnamefont {S.}~\bibnamefont
  {Sukhov}}\ and\ \bibinfo {author} {\bibfnamefont {A.}~\bibnamefont
  {Dogariu}},\ }\bibfield  {title} {\enquote {\bibinfo {title}
  {Non-conservative optical forces},}\ }\href@noop {} {\bibfield  {journal}
  {\bibinfo  {journal} {Rep. Prog. Phys.}\ }\textbf {\bibinfo {volume} {80}},\
  \bibinfo {pages} {112001} (\bibinfo {year} {2017})}\BibitemShut {NoStop}%
\bibitem [{\citenamefont {Stokes}(1847)}]{Stokes}%
  \BibitemOpen
  \bibfield  {author} {\bibinfo {author} {\bibfnamefont {G.~G.}\ \bibnamefont
  {Stokes}},\ }\bibfield  {title} {\enquote {\bibinfo {title} {On the theory of
  oscillatory waves},}\ }\href@noop {} {\bibfield  {journal} {\bibinfo
  {journal} {Trans. Cambridge Philos. Soc.}\ }\textbf {\bibinfo {volume} {8}},\
  \bibinfo {pages} {441--455} (\bibinfo {year} {1847})}\BibitemShut {NoStop}%
\bibitem [{\citenamefont {van~den Bremer}\ and\ \citenamefont
  {Breivik}(2017)}]{Bremer2017}%
  \BibitemOpen
  \bibfield  {author} {\bibinfo {author} {\bibfnamefont {T.~S.}\ \bibnamefont
  {van~den Bremer}}\ and\ \bibinfo {author} {\bibfnamefont {{\O}.}~\bibnamefont
  {Breivik}},\ }\bibfield  {title} {\enquote {\bibinfo {title} {Stokes
  drift},}\ }\href@noop {} {\bibfield  {journal} {\bibinfo  {journal} {Philos.
  Trans. R. Soc. A}\ }\textbf {\bibinfo {volume} {376}},\ \bibinfo {pages}
  {20170104} (\bibinfo {year} {2017})}\BibitemShut {NoStop}%
\bibitem [{\citenamefont {Falkovich}(2018)}]{Falkovich_book}%
  \BibitemOpen
  \bibfield  {author} {\bibinfo {author} {\bibfnamefont {G.}~\bibnamefont
  {Falkovich}},\ }\href@noop {} {\emph {\bibinfo {title} {Fluid Mechanics}}},\
  \bibinfo {edition} {2nd}\ ed.\ (\bibinfo  {publisher} {Cambridge University
  Press},\ \bibinfo {year} {2018})\BibitemShut {NoStop}%
\bibitem [{\citenamefont {Gaponov}\ and\ \citenamefont
  {Miller}(1958)}]{Gaponov1958}%
  \BibitemOpen
  \bibfield  {author} {\bibinfo {author} {\bibfnamefont {A.~V.}\ \bibnamefont
  {Gaponov}}\ and\ \bibinfo {author} {\bibfnamefont {M.~A.}\ \bibnamefont
  {Miller}},\ }\bibfield  {title} {\enquote {\bibinfo {title} {Potential wells
  for charged particles in a high-frequency electromagnetic field},}\
  }\href@noop {} {\bibfield  {journal} {\bibinfo  {journal} {Sov. Phys. JETP}\
  }\textbf {\bibinfo {volume} {7}},\ \bibinfo {pages} {168--169} (\bibinfo
  {year} {1958})}\BibitemShut {NoStop}%
\bibitem [{\citenamefont {Gordon}(1973)}]{Gordon1973}%
  \BibitemOpen
  \bibfield  {author} {\bibinfo {author} {\bibfnamefont {J.~P.}\ \bibnamefont
  {Gordon}},\ }\bibfield  {title} {\enquote {\bibinfo {title} {Radiation forces
  and momenta in dielectric media},}\ }\href@noop {} {\bibfield  {journal}
  {\bibinfo  {journal} {Phys. Rev. A}\ }\textbf {\bibinfo {volume} {8}},\
  \bibinfo {pages} {14} (\bibinfo {year} {1973})}\BibitemShut {NoStop}%
\bibitem [{\citenamefont {Washimi}\ and\ \citenamefont
  {Karpman}(1976)}]{Washimi1976}%
  \BibitemOpen
  \bibfield  {author} {\bibinfo {author} {\bibfnamefont {H.}~\bibnamefont
  {Washimi}}\ and\ \bibinfo {author} {\bibfnamefont {V.~I.}\ \bibnamefont
  {Karpman}},\ }\bibfield  {title} {\enquote {\bibinfo {title} {The
  ponderomotive force of a high-frequency electromagnetic field in a dispersive
  medium},}\ }\href@noop {} {\bibfield  {journal} {\bibinfo  {journal} {Sov.
  Phys. JETP}\ }\textbf {\bibinfo {volume} {44}},\ \bibinfo {pages} {528--531}
  (\bibinfo {year} {1976})}\BibitemShut {NoStop}%
\bibitem [{\citenamefont {Bliokh}\ \emph {et~al.}(2017)\citenamefont {Bliokh},
  \citenamefont {Bekshaev},\ and\ \citenamefont {Nori}}]{Bliokh2017NJP}%
  \BibitemOpen
  \bibfield  {author} {\bibinfo {author} {\bibfnamefont {K.~Y.}\ \bibnamefont
  {Bliokh}}, \bibinfo {author} {\bibfnamefont {A.~Y.}\ \bibnamefont
  {Bekshaev}}, \ and\ \bibinfo {author} {\bibfnamefont {F.}~\bibnamefont
  {Nori}},\ }\bibfield  {title} {\enquote {\bibinfo {title} {{Optical momentum
  and angular momentum in complex media: from the Abraham--Minkowski debate to
  unusual properties of surface plasmon-polaritons}},}\ }\href@noop {}
  {\bibfield  {journal} {\bibinfo  {journal} {New J. Phys.}\ }\textbf {\bibinfo
  {volume} {19}},\ \bibinfo {pages} {123014} (\bibinfo {year}
  {2017})}\BibitemShut {NoStop}%
\bibitem [{\citenamefont {Jones}\ and\ \citenamefont
  {Leslie}(1978)}]{Jones1978}%
  \BibitemOpen
  \bibfield  {author} {\bibinfo {author} {\bibfnamefont {R.~V.}\ \bibnamefont
  {Jones}}\ and\ \bibinfo {author} {\bibfnamefont {B.}~\bibnamefont {Leslie}},\
  }\bibfield  {title} {\enquote {\bibinfo {title} {The measurement of optical
  radiation pressure in dispersive media},}\ }\href@noop {} {\bibfield
  {journal} {\bibinfo  {journal} {{Proc. R. Soc. Lond. A}}\ }\textbf {\bibinfo
  {volume} {360}},\ \bibinfo {pages} {347} (\bibinfo {year}
  {1978})}\BibitemShut {NoStop}%
\bibitem [{\citenamefont {Campbell}\ \emph {et~al.}(2005)\citenamefont
  {Campbell}, \citenamefont {Leanhardt}, \citenamefont {Mun}, \citenamefont
  {Boyd}, \citenamefont {Streed}, \citenamefont {Ketterle},\ and\ \citenamefont
  {Pritchard}}]{Campbell2005}%
  \BibitemOpen
  \bibfield  {author} {\bibinfo {author} {\bibfnamefont {G.~K.}\ \bibnamefont
  {Campbell}}, \bibinfo {author} {\bibfnamefont {A.~E.}\ \bibnamefont
  {Leanhardt}}, \bibinfo {author} {\bibfnamefont {J.}~\bibnamefont {Mun}},
  \bibinfo {author} {\bibfnamefont {M.}~\bibnamefont {Boyd}}, \bibinfo {author}
  {\bibfnamefont {E.~W.}\ \bibnamefont {Streed}}, \bibinfo {author}
  {\bibfnamefont {W.}~\bibnamefont {Ketterle}}, \ and\ \bibinfo {author}
  {\bibfnamefont {D.~E.}\ \bibnamefont {Pritchard}},\ }\bibfield  {title}
  {\enquote {\bibinfo {title} {Photon recoil momentum in dispersive media},}\
  }\href@noop {} {\bibfield  {journal} {\bibinfo  {journal} {Phys. Rev. Lett.}\
  }\textbf {\bibinfo {volume} {94}},\ \bibinfo {pages} {170403} (\bibinfo
  {year} {2005})}\BibitemShut {NoStop}%
\bibitem [{\citenamefont {Astrath}\ \emph {et~al.}(2022)\citenamefont
  {Astrath}, \citenamefont {Flizikowski}, \citenamefont {Anghinoni},
  \citenamefont {Malacarne}, \citenamefont {Baesso}, \citenamefont {Po{\v
  z}ar}, \citenamefont {Partanen}, \citenamefont {Brevik}, \citenamefont
  {Razansky},\ and\ \citenamefont {Bialkowski}}]{Astrath2022}%
  \BibitemOpen
  \bibfield  {author} {\bibinfo {author} {\bibfnamefont {N.~G.~C.}\
  \bibnamefont {Astrath}}, \bibinfo {author} {\bibfnamefont {G.~A.~S.}\
  \bibnamefont {Flizikowski}}, \bibinfo {author} {\bibfnamefont
  {B.}~\bibnamefont {Anghinoni}}, \bibinfo {author} {\bibfnamefont {L.~C.}\
  \bibnamefont {Malacarne}}, \bibinfo {author} {\bibfnamefont {M.~L.}\
  \bibnamefont {Baesso}}, \bibinfo {author} {\bibfnamefont {T.}~\bibnamefont
  {Po{\v z}ar}}, \bibinfo {author} {\bibfnamefont {M.}~\bibnamefont
  {Partanen}}, \bibinfo {author} {\bibfnamefont {I.}~\bibnamefont {Brevik}},
  \bibinfo {author} {\bibfnamefont {D.}~\bibnamefont {Razansky}}, \ and\
  \bibinfo {author} {\bibfnamefont {S.~E.}\ \bibnamefont {Bialkowski}},\
  }\bibfield  {title} {\enquote {\bibinfo {title} {Unveiling bulk and surface
  radiation forces in a dielectric liquid},}\ }\href@noop {} {\bibfield
  {journal} {\bibinfo  {journal} {Light: Science \& Applications}\ }\textbf
  {\bibinfo {volume} {11}},\ \bibinfo {pages} {103} (\bibinfo {year}
  {2022})}\BibitemShut {NoStop}%
\bibitem [{\citenamefont {Berry}(2009)}]{Berry2009}%
  \BibitemOpen
  \bibfield  {author} {\bibinfo {author} {\bibfnamefont {M.~V.}\ \bibnamefont
  {Berry}},\ }\bibfield  {title} {\enquote {\bibinfo {title} {Optical
  currents},}\ }\href@noop {} {\bibfield  {journal} {\bibinfo  {journal} {J.
  Opt. A: Pure Appl. Opt.}\ }\textbf {\bibinfo {volume} {11}},\ \bibinfo
  {pages} {094001} (\bibinfo {year} {2009})}\BibitemShut {NoStop}%
\bibitem [{\citenamefont {Bliokh}\ \emph
  {et~al.}(2014{\natexlab{b}})\citenamefont {Bliokh}, \citenamefont
  {Bekshaev},\ and\ \citenamefont {Nori}}]{Bliokh2014NC}%
  \BibitemOpen
  \bibfield  {author} {\bibinfo {author} {\bibfnamefont {K.~Y.}\ \bibnamefont
  {Bliokh}}, \bibinfo {author} {\bibfnamefont {A.~Y.}\ \bibnamefont
  {Bekshaev}}, \ and\ \bibinfo {author} {\bibfnamefont {F.}~\bibnamefont
  {Nori}},\ }\bibfield  {title} {\enquote {\bibinfo {title} {Extraordinary
  momentum and spin in evanescent waves},}\ }\href@noop {} {\bibfield
  {journal} {\bibinfo  {journal} {Nat. Commun.}\ }\textbf {\bibinfo {volume}
  {5}},\ \bibinfo {pages} {3300} (\bibinfo {year}
  {2014}{\natexlab{b}})}\BibitemShut {NoStop}%
\bibitem [{\citenamefont {Bliokh}\ and\ \citenamefont
  {Nori}(2015)}]{Bliokh2015PR}%
  \BibitemOpen
  \bibfield  {author} {\bibinfo {author} {\bibfnamefont {K.~Y.}\ \bibnamefont
  {Bliokh}}\ and\ \bibinfo {author} {\bibfnamefont {F.}~\bibnamefont {Nori}},\
  }\bibfield  {title} {\enquote {\bibinfo {title} {Transverse and longitudinal
  angular momenta of light},}\ }\href@noop {} {\bibfield  {journal} {\bibinfo
  {journal} {Phys. Rep.}\ }\textbf {\bibinfo {volume} {592}},\ \bibinfo {pages}
  {1--38} (\bibinfo {year} {2015})}\BibitemShut {NoStop}%
\bibitem [{\citenamefont {Shi}\ \emph {et~al.}(2019)\citenamefont {Shi},
  \citenamefont {Zhao}, \citenamefont {Long}, \citenamefont {Yang},
  \citenamefont {Wang}, \citenamefont {Chen}, \citenamefont {Ren},\ and\
  \citenamefont {Zhang}}]{Shi2019}%
  \BibitemOpen
  \bibfield  {author} {\bibinfo {author} {\bibfnamefont {C.}~\bibnamefont
  {Shi}}, \bibinfo {author} {\bibfnamefont {R.}~\bibnamefont {Zhao}}, \bibinfo
  {author} {\bibfnamefont {Y.}~\bibnamefont {Long}}, \bibinfo {author}
  {\bibfnamefont {S.}~\bibnamefont {Yang}}, \bibinfo {author} {\bibfnamefont
  {Y.}~\bibnamefont {Wang}}, \bibinfo {author} {\bibfnamefont {H.}~\bibnamefont
  {Chen}}, \bibinfo {author} {\bibfnamefont {J.}~\bibnamefont {Ren}}, \ and\
  \bibinfo {author} {\bibfnamefont {X.}~\bibnamefont {Zhang}},\ }\bibfield
  {title} {\enquote {\bibinfo {title} {Observation of acoustic spin},}\
  }\href@noop {} {\bibfield  {journal} {\bibinfo  {journal} {Natl. Sci. Rev.}\
  }\textbf {\bibinfo {volume} {6}},\ \bibinfo {pages} {707} (\bibinfo {year}
  {2019})}\BibitemShut {NoStop}%
\bibitem [{\citenamefont {Burns}\ \emph {et~al.}(2020)\citenamefont {Burns},
  \citenamefont {Bliokh}, \citenamefont {Nori},\ and\ \citenamefont
  {Dressel}}]{Burns2020}%
  \BibitemOpen
  \bibfield  {author} {\bibinfo {author} {\bibfnamefont {L.}~\bibnamefont
  {Burns}}, \bibinfo {author} {\bibfnamefont {K.~Y.}\ \bibnamefont {Bliokh}},
  \bibinfo {author} {\bibfnamefont {F.}~\bibnamefont {Nori}}, \ and\ \bibinfo
  {author} {\bibfnamefont {J.}~\bibnamefont {Dressel}},\ }\bibfield  {title}
  {\enquote {\bibinfo {title} {Acoustic versus electromagnetic field theory:
  scalar, vector, spinor representations and the emergence of acoustic spin},}\
  }\href@noop {} {\bibfield  {journal} {\bibinfo  {journal} {New J. Phys.}\
  }\textbf {\bibinfo {volume} {22}},\ \bibinfo {pages} {053050} (\bibinfo
  {year} {2020})}\BibitemShut {NoStop}%
\bibitem [{\citenamefont {Bliokh}\ \emph {et~al.}(2022)\citenamefont {Bliokh},
  \citenamefont {Punzmann}, \citenamefont {Xia}, \citenamefont {Nori},\ and\
  \citenamefont {Shats}}]{Bliokh2022}%
  \BibitemOpen
  \bibfield  {author} {\bibinfo {author} {\bibfnamefont {K.~Y.}\ \bibnamefont
  {Bliokh}}, \bibinfo {author} {\bibfnamefont {H.}~\bibnamefont {Punzmann}},
  \bibinfo {author} {\bibfnamefont {H.}~\bibnamefont {Xia}}, \bibinfo {author}
  {\bibfnamefont {F.}~\bibnamefont {Nori}}, \ and\ \bibinfo {author}
  {\bibfnamefont {M.}~\bibnamefont {Shats}},\ }\bibfield  {title} {\enquote
  {\bibinfo {title} {Field theory spin and momentum in water waves},}\
  }\href@noop {} {\bibfield  {journal} {\bibinfo  {journal} {Sci. Adv.}\
  }\textbf {\bibinfo {volume} {8}},\ \bibinfo {pages} {eabm1295} (\bibinfo
  {year} {2022})}\BibitemShut {NoStop}%
\bibitem [{\citenamefont {Bliokh}\ and\ \citenamefont
  {Bliokh}(2022)}]{Bliokh2022_II}%
  \BibitemOpen
  \bibfield  {author} {\bibinfo {author} {\bibfnamefont {K.~Y.}\ \bibnamefont
  {Bliokh}}\ and\ \bibinfo {author} {\bibfnamefont {Y.~P.}\ \bibnamefont
  {Bliokh}},\ }\bibfield  {title} {\enquote {\bibinfo {title} {Momentum,
  angular momentum, and spin of waves in plasma},}\ }\href@noop {} {\bibfield
  {journal} {\bibinfo  {journal} {arXiv:2203.05240}\ } (\bibinfo {year}
  {2022})}\BibitemShut {NoStop}%
\bibitem [{\citenamefont {Dodin}\ and\ \citenamefont
  {Fisch}(2012)}]{Dodin2012}%
  \BibitemOpen
  \bibfield  {author} {\bibinfo {author} {\bibfnamefont {I.~Y.}\ \bibnamefont
  {Dodin}}\ and\ \bibinfo {author} {\bibfnamefont {N.~J.}\ \bibnamefont
  {Fisch}},\ }\bibfield  {title} {\enquote {\bibinfo {title} {{Axiomatic
  geometrical optics, Abraham-Minkowski controversy, and photon properties
  derived classically}},}\ }\href@noop {} {\bibfield  {journal} {\bibinfo
  {journal} {Phys. Rev. A}\ }\textbf {\bibinfo {volume} {86}},\ \bibinfo
  {pages} {053834} (\bibinfo {year} {2012})}\BibitemShut {NoStop}%
\bibitem [{\citenamefont {Krall}\ and\ \citenamefont
  {Trivelpiece}(1973)}]{Krall}%
  \BibitemOpen
  \bibfield  {author} {\bibinfo {author} {\bibfnamefont {N.~A.}\ \bibnamefont
  {Krall}}\ and\ \bibinfo {author} {\bibfnamefont {A.~W.}\ \bibnamefont
  {Trivelpiece}},\ }\href@noop {} {\emph {\bibinfo {title} {Principles of
  Plasma Physics}}}\ (\bibinfo  {publisher} {McGraw-Hill, New York},\ \bibinfo
  {year} {1973})\BibitemShut {NoStop}%
\bibitem [{\citenamefont {Stenholm}(1986)}]{Stenholm1986}%
  \BibitemOpen
  \bibfield  {author} {\bibinfo {author} {\bibfnamefont {S.}~\bibnamefont
  {Stenholm}},\ }\bibfield  {title} {\enquote {\bibinfo {title} {The
  semiclassical theory of laser cooling},}\ }\href@noop {} {\bibfield
  {journal} {\bibinfo  {journal} {Rev. Mod. Phys.}\ }\textbf {\bibinfo {volume}
  {58}},\ \bibinfo {pages} {699} (\bibinfo {year} {1986})}\BibitemShut
  {NoStop}%
\bibitem [{\citenamefont {Chu}(1998)}]{Chu1998}%
  \BibitemOpen
  \bibfield  {author} {\bibinfo {author} {\bibfnamefont {S.}~\bibnamefont
  {Chu}},\ }\bibfield  {title} {\enquote {\bibinfo {title} {The manipulation of
  neutral particles},}\ }\href@noop {} {\bibfield  {journal} {\bibinfo
  {journal} {Rev. Mod. Phys.}\ }\textbf {\bibinfo {volume} {70}},\ \bibinfo
  {pages} {685} (\bibinfo {year} {1998})}\BibitemShut {NoStop}%
\bibitem [{\citenamefont {Phillips}(1998)}]{Phillips1998}%
  \BibitemOpen
  \bibfield  {author} {\bibinfo {author} {\bibfnamefont {W.~D.}\ \bibnamefont
  {Phillips}},\ }\bibfield  {title} {\enquote {\bibinfo {title} {Laser cooling
  and trapping of neutral atoms},}\ }\href@noop {} {\bibfield  {journal}
  {\bibinfo  {journal} {Rev. Mod. Phys.}\ }\textbf {\bibinfo {volume} {70}},\
  \bibinfo {pages} {721} (\bibinfo {year} {1998})}\BibitemShut {NoStop}%
\bibitem [{\citenamefont {Aspelmeyer}\ \emph {et~al.}(2014)\citenamefont
  {Aspelmeyer}, \citenamefont {Kippenberg},\ and\ \citenamefont
  {Marquardt}}]{Aspelmeyer2014}%
  \BibitemOpen
  \bibfield  {author} {\bibinfo {author} {\bibfnamefont {M.}~\bibnamefont
  {Aspelmeyer}}, \bibinfo {author} {\bibfnamefont {T.~J.}\ \bibnamefont
  {Kippenberg}}, \ and\ \bibinfo {author} {\bibfnamefont {F.}~\bibnamefont
  {Marquardt}},\ }\bibfield  {title} {\enquote {\bibinfo {title} {Cavity
  optomechanics},}\ }\href@noop {} {\bibfield  {journal} {\bibinfo  {journal}
  {Rev. Mod. Phys.}\ }\textbf {\bibinfo {volume} {86}},\ \bibinfo {pages}
  {1391} (\bibinfo {year} {2014})}\BibitemShut {NoStop}%
\bibitem [{\citenamefont {Ding}\ \emph {et~al.}(2013)\citenamefont {Ding},
  \citenamefont {Li}, \citenamefont {Lin}, \citenamefont {Stratton},
  \citenamefont {Nama}, \citenamefont {Guo}, \citenamefont {Slotcavage},
  \citenamefont {Mao}, \citenamefont {Shi}, \citenamefont {Costanzo},\ and\
  \citenamefont {Huang}}]{Ding2013}%
  \BibitemOpen
  \bibfield  {author} {\bibinfo {author} {\bibfnamefont {X.}~\bibnamefont
  {Ding}}, \bibinfo {author} {\bibfnamefont {P.}~\bibnamefont {Li}}, \bibinfo
  {author} {\bibfnamefont {S.-C.~S.}\ \bibnamefont {Lin}}, \bibinfo {author}
  {\bibfnamefont {Z.~S.}\ \bibnamefont {Stratton}}, \bibinfo {author}
  {\bibfnamefont {N.}~\bibnamefont {Nama}}, \bibinfo {author} {\bibfnamefont
  {F.}~\bibnamefont {Guo}}, \bibinfo {author} {\bibfnamefont {D.}~\bibnamefont
  {Slotcavage}}, \bibinfo {author} {\bibfnamefont {X.}~\bibnamefont {Mao}},
  \bibinfo {author} {\bibfnamefont {J.}~\bibnamefont {Shi}}, \bibinfo {author}
  {\bibfnamefont {F.}~\bibnamefont {Costanzo}}, \ and\ \bibinfo {author}
  {\bibfnamefont {T.~J.}\ \bibnamefont {Huang}},\ }\bibfield  {title} {\enquote
  {\bibinfo {title} {Surface acoustic wave microfluidics},}\ }\href@noop {}
  {\bibfield  {journal} {\bibinfo  {journal} {Lab Chip.}\ }\textbf {\bibinfo
  {volume} {13}},\ \bibinfo {pages} {3626} (\bibinfo {year}
  {2013})}\BibitemShut {NoStop}%
\bibitem [{\citenamefont {Landau}\ and\ \citenamefont
  {Lifshitz}(1987)}]{LLfluid}%
  \BibitemOpen
  \bibfield  {author} {\bibinfo {author} {\bibfnamefont {L.~D.}\ \bibnamefont
  {Landau}}\ and\ \bibinfo {author} {\bibfnamefont {E.~M.}\ \bibnamefont
  {Lifshitz}},\ }\href@noop {} {\emph {\bibinfo {title} {Fluid Mechanics}}}\
  (\bibinfo  {publisher} {{Butterworth-Heinemann, Oxford}},\ \bibinfo {year}
  {1987})\BibitemShut {NoStop}%
\bibitem [{\citenamefont {Maier}(2007)}]{Maier_book}%
  \BibitemOpen
  \bibfield  {author} {\bibinfo {author} {\bibfnamefont {S.~A.}\ \bibnamefont
  {Maier}},\ }\href@noop {} {\emph {\bibinfo {title} {Plasmonics: Fundamentals
  and Applications}}}\ (\bibinfo  {publisher} {Springer, Berlin},\ \bibinfo
  {year} {2007})\BibitemShut {NoStop}%
\bibitem [{\citenamefont {Einstein}\ and\ \citenamefont
  {Laub}(1908)}]{Einstein1908}%
  \BibitemOpen
  \bibfield  {author} {\bibinfo {author} {\bibfnamefont {A.}~\bibnamefont
  {Einstein}}\ and\ \bibinfo {author} {\bibfnamefont {J.}~\bibnamefont
  {Laub}},\ }\bibfield  {title} {\enquote {\bibinfo {title} {{\"{U}ber die im
  elektromagnetischen Felde auf ruhende K\"{o}rper ausge\"{u}bten
  ponderomotorischen Kr\"{a}fte}},}\ }\href@noop {} {\bibfield  {journal}
  {\bibinfo  {journal} {Annalen der Physik}\ }\textbf {\bibinfo {volume}
  {331}},\ \bibinfo {pages} {541--550} (\bibinfo {year} {1908})}\BibitemShut
  {NoStop}%
\bibitem [{\citenamefont {Mansuripur}(2014)}]{Mansuripur2014}%
  \BibitemOpen
  \bibfield  {author} {\bibinfo {author} {\bibfnamefont {M.}~\bibnamefont
  {Mansuripur}},\ }\bibfield  {title} {\enquote {\bibinfo {title} {{The Lorentz
  Force Law and its Connections to Hidden Momentum, the Einstein-Laub Force,
  and the Aharonov-Casher Effect}},}\ }\href@noop {} {\bibfield  {journal}
  {\bibinfo  {journal} {IEEE Trans. Magn.}\ }\textbf {\bibinfo {volume} {50}},\
  \bibinfo {pages} {1300110} (\bibinfo {year} {2014})}\BibitemShut {NoStop}%
\bibitem [{\citenamefont {Webb}(2016)}]{Webb2016}%
  \BibitemOpen
  \bibfield  {author} {\bibinfo {author} {\bibfnamefont {K.~J.}\ \bibnamefont
  {Webb}},\ }\bibfield  {title} {\enquote {\bibinfo {title} {{Relationship
  between the Einstein-Laub electromagnetic force and the Lorentz force on free
  charge}},}\ }\href@noop {} {\bibfield  {journal} {\bibinfo  {journal} {Phys.
  Rev. B}\ }\textbf {\bibinfo {volume} {94}},\ \bibinfo {pages} {064203}
  (\bibinfo {year} {2016})}\BibitemShut {NoStop}%
\bibitem [{\citenamefont {Philbin}(2011)}]{Philbin2011}%
  \BibitemOpen
  \bibfield  {author} {\bibinfo {author} {\bibfnamefont {T.~G.}\ \bibnamefont
  {Philbin}},\ }\bibfield  {title} {\enquote {\bibinfo {title} {Electromagnetic
  energy momentum in dispersive media},}\ }\href@noop {} {\bibfield  {journal}
  {\bibinfo  {journal} {Phys. Rev. A}\ }\textbf {\bibinfo {volume} {83}},\
  \bibinfo {pages} {013823} (\bibinfo {year} {2011})}\BibitemShut {NoStop}%
\bibitem [{\citenamefont {Philbin}(2012)}]{Philbin2012}%
  \BibitemOpen
  \bibfield  {author} {\bibinfo {author} {\bibfnamefont {T.~G.}\ \bibnamefont
  {Philbin}},\ }\bibfield  {title} {\enquote {\bibinfo {title} {{Erratum:
  Electromagnetic energy momentum in dispersive media [Phys. Rev. A 83, 013823
  (2011)]}},}\ }\href@noop {} {\bibfield  {journal} {\bibinfo  {journal} {Phys.
  Rev. A}\ }\textbf {\bibinfo {volume} {85}},\ \bibinfo {pages} {059902(E)}
  (\bibinfo {year} {2012})}\BibitemShut {NoStop}%
\bibitem [{\citenamefont {Nelson}(1991)}]{Nelson1991}%
  \BibitemOpen
  \bibfield  {author} {\bibinfo {author} {\bibfnamefont {D.~F.}\ \bibnamefont
  {Nelson}},\ }\bibfield  {title} {\enquote {\bibinfo {title} {{Momentum,
  pseudomomentum, and wave momentum: Toward resolving the Minkowski-Abraham
  controversy}},}\ }\href@noop {} {\bibfield  {journal} {\bibinfo  {journal}
  {Phys. Rev. A}\ }\textbf {\bibinfo {volume} {44}},\ \bibinfo {pages}
  {3985--3996} (\bibinfo {year} {1991})}\BibitemShut {NoStop}%
\bibitem [{\citenamefont {Barnett}(2010)}]{Barnett2010}%
  \BibitemOpen
  \bibfield  {author} {\bibinfo {author} {\bibfnamefont {S.~M.}\ \bibnamefont
  {Barnett}},\ }\bibfield  {title} {\enquote {\bibinfo {title} {Rotation of
  electromagnetic fields and the nature of optical angular momentum},}\
  }\href@noop {} {\bibfield  {journal} {\bibinfo  {journal} {J. Mod. Opt.}\
  }\textbf {\bibinfo {volume} {57}},\ \bibinfo {pages} {1339--1343} (\bibinfo
  {year} {2010})}\BibitemShut {NoStop}%
\bibitem [{\citenamefont {Partanen}\ and\ \citenamefont
  {Tulkki}(2017)}]{Partanen2017}%
  \BibitemOpen
  \bibfield  {author} {\bibinfo {author} {\bibfnamefont {M.}~\bibnamefont
  {Partanen}}\ and\ \bibinfo {author} {\bibfnamefont {J.}~\bibnamefont
  {Tulkki}},\ }\bibfield  {title} {\enquote {\bibinfo {title} {Mass-polariton
  theory of light in dispersive media},}\ }\href@noop {} {\bibfield  {journal}
  {\bibinfo  {journal} {Phys. Rev. A}\ }\textbf {\bibinfo {volume} {96}},\
  \bibinfo {pages} {063834} (\bibinfo {year} {2017})}\BibitemShut {NoStop}%
\end{thebibliography}%

\end{document}